\documentclass[12pt]{article}
\usepackage{geometry}
\usepackage{latexsym}
\usepackage{amssymb}
\usepackage{amsmath}
\usepackage{multirow}
\usepackage{enumitem}

\begin{document}

\newtheorem{thm}{Theorem}[section]
\newtheorem{lemm}[thm]{Lemma}
\newtheorem{prop}[thm]{Proposition}
\newtheorem{defn}[thm]{Definition}
\newtheorem{coro}[thm]{Corollary}
\newtheorem{algo}[thm]{Algorithm}
\newtheorem{proc}[thm]{Procedure}
\newtheorem{conj}[thm]{Conjecture}
\newtheorem{para}[thm]{Paradox}
\newtheorem{examp}[thm]{Example}
\newenvironment{exmp}{\begin{examp}\rm}{\hspace{.1in}\hfill{$\square$}\end{examp}}

\newcommand{\ignore}[1]{}
\newcommand{\Proof}{\noindent{\sl Proof}.\ } 
\newcommand{\Endproof}{{\hfill{$\square$}\vspace{.1in}}}
\newcommand{\orthogonal}{{+}}
\newcommand{\T}{\top}
\newcommand{\F}{\bot}
\newcommand{\E}{\emptyset}
\newcommand{\Iff}{if and only if}
\newcommand{\OR}{\vee}
\newcommand{\AND}{\wedge}
\newcommand{\XOR}{\oplus}
\newcommand{\NOR}{\downarrow}
\newcommand{\NAND}{\uparrow}
\newcommand{\IFF}{\leftrightarrow}
\newcommand{\ssub}{-}
\newcommand{\hs}[1]{\hspace{#1}}
\newcommand{\tuple}[1]{\langle #1 \rangle}
\newcommand{\simply}{\Longrightarrow}
\newcommand{\rew}{\rightarrow}
\newcommand{\Rew}{\Rightarrow}
\newcommand{\imply}{\rightarrow}
\newcommand{\toe}{\approx}
\newcommand{\tog}{\succ}
\newcommand{\togm}{\succ^{mul}}
\newcommand{\togl}{\succ^{lex}}
\newcommand{\toge}{\succeq}
\newcommand{\togem}{\succeq^{mul}}
\newcommand{\togel}{\succeq^{lex}}

\newcommand{\subst}{\leftharpoonup}
\newcommand{\assign}{\mapsto}
\newcommand{\entail}{\models}
\newcommand{\sateq}{\approx}
\newcommand{\infer}{\vdash}
\newcommand{\move}{\vdash}
\newcommand{\inferC}[1]{\vdash_{\cal #1}}
\newcommand{\sequ}{\Rightarrow}
\newcommand{\func}{\mapsto}
\newcommand{\ov}{\overline}
\newcommand{\BB}{{\it HyMaxSATBB}}
\newcommand{\search}{\mbox{\it engine}}
\newcommand{\lunify}{\mbox{\it lunify}}
\newcommand{\unify}{\mbox{\it unify}}
\newcommand{\eval}{\mbox{\it eval}}
\newcommand{\DPLL}{{\it DPLL}}
\newcommand{\DPLLT}{{\it DPLLT}}
\newcommand{\GSAT}{{\it GSAT}}
\newcommand{\IF}{{\bf if}\ }
\newcommand{\FI}{{\bf fi}}
\newcommand{\THEN}{{\bf then}\ }
\newcommand{\ELIF}{{\bf else~if}\ }
\newcommand{\ELSE}{{\bf else}\ }
\newcommand{\END}{{\bf end}}
\newcommand{\WHILE}{{\bf while}\ }
\newcommand{\RETURN}{{\bf return}\ }
\newcommand{\PROC}{{\bf proc}\ }
\newcommand{\DO}{{\bf do}\ }
\newcommand{\OD}{{\bf od}}
\newcommand{\FOR}{{\bf for}\ }
\newcommand{\TO}{{\bf to}\ }
\newcommand{\etal}{{\it et al.}}
\newcommand{\Union}{\mbox{\it Union}}
\newcommand{\Find}{\mbox{\it Find}}
\newcommand{\nf}{\mbox{\it nf}}
\newcommand{\NF}{\mbox{\it NF}}

\newcommand{\bfx}{{\ensuremath{\mathbf{x}}}}
\newcommand{\thetadd}{{\ensuremath{\bar{\theta}^{(\cdot)}_\cdot}}}

\newcommand{\smsq}{\mbox{\scriptsize $\square$}}
\newcommand{\U}{\mbox{\scriptsize$\cal \,U\,$}}
\newcommand{\R}{\mbox{\scriptsize$\cal \,R\,$}}
\newcommand{\W}{\mbox{\scriptsize$\cal \,W\,$}}
\newcommand{\SR}{\mbox{\scriptsize$\cal \,S\,$}}

\newcommand{\B}{{\cal B}}
\newcommand{\M}{{\cal M}}
\newcommand{\N}{{\cal N}}
\newcommand{\Z}{{\cal Z}}
\newcommand{\CE}{{\cal E}}
\newcommand{\CF}{{\cal F}}
\newcommand{\CP}{{\cal P}}
\newcommand{\CR}{{\cal R}}
\newcommand{\CT}{{\cal T}}

\newcommand{\Godel}{{G\"{o}del}}
\newcommand{\Erdos}{{Erd\"{o}s}}

\title{Countability versus Computability\footnote
{DOI: 10.13140/RG.2.2.31203.92961}}

\author{
\bf Hantao Zhang \\
\em Department of Computer Science \\
\em The University of Iowa \\
\em Iowa City, Iowa, USA \\
\em hantao-zhang@uiowa.edu
}
\date{September 29, 2026}

\maketitle

\ignore{
\begin{abstract}
  The concept of {\em countable sets} is attributed to Georg Cantor, 
  who established the distinction between countable and uncountable sets in 1874. 
  The concept of {\em computable sets} emerged in the 1930s 
  through the foundational work on computing models by \Godel, Church, and Turing. 
  In this paper, we investigate the connection between countability and computability.
  A {\em counting bijection} of a set $S$ is a
  bijection from the set of natural numbers to $S$.  We say $S$
  is {\em enumerable} if it is either finite or admitting a computable
  counting bijection.  Our initial investigation shows that a set $S$ is enumerable if and
  only if it is computable.  This equivalence offers new insights into
  set theory and computability theory. 
  
  We further show that a set is countable if and only if
  it admits a {\em counting order}, which is a well-order satisfying
  the {\em nearby} property.  Based on this property,
  we provide a procedure whose existence gives a 
  necessary and sufficient condition for the countability of a set. 
  This procedure is an algorithm if and only if the set is computable. 
  
  The {\em characteristic function} of a set is a total Boolean function which tells if
  an element is a member of the set or not. Given
  an infinite set $S$ of natural numbers, we prove the coexistences
  of a counting bijection of $S$ and its characteristic function in first-order arithmetic. 
  This result has a significant implication: the standard proof 
  that every set of natural numbers is countable is invalid. This is because
  ``every set $S$ of natural numbers is countable'' implies ``a 
  counting bijection of $S$ is definable in first-order arithmetic,'' 
  which (by our result) would imply that a characteristic function of $S$ is definable in first-order 
  arithmetic. This contradicts the fact that there are
  uncountably many sets of natural numbers whose characteristic functions
  are undefinable in first-order arithmetic. 
\end{abstract}}

\begin{abstract}
  We investigate the connection between countability and computability.
  It is known that a set computable if and only if it admits a computable counting bijection
  and that a set is decidable if and only if it admits an increasing computable counting bijection.
  We further show that a set is countable if and only if
  it admits a {\em counting order}, which is a well-order satisfying
  the {\em nearby} property.  Based on this property,
  we provide a procedure whose existence gives a 
  necessary and sufficient condition for the countability of a set. 
  This procedure is an algorithm if and only if the set is computable. 
  Given an infinite set $S$ of natural numbers, we prove the coexistence
  of a counting bijection of $S$ and its characteristic function in first-order arithmetic. 
  In the process of proving this result, we found that the standard proof 
  that every set of natural numbers is countable is invalid. 
\end{abstract}

\section{Introduction}

In 1874, Georg Cantor published the first proof that
there is no one-to-one correspondence between the set of all real
numbers (which is uncountable) and the set of all natural numbers
(which is countable)  \cite{Cantor}.  In 1891, Cantor simplified his proof using the
well-known diagonal argument, which has since found important applications in 
both set theory and computability theory \cite{Sipser}.

In the 1930s, several independent attempts were made to formalize the
notion of computability: 
Kurt \Godel's partial recursive functions \cite{Godel}, 
Alonzo Church's lambda calculus \cite{Church}, and 
Turing's formal model, later known as Turing machines  \cite{Turing}.
Church and Turing proved that these three notions of
computable functions coincide and proposed the well-known conjecture
called the {\em Church-Turing thesis}: A function is computable if it can be computed
by a Turing machine \cite{Cop}.  Other formal attempts to characterize computability,
including Kleene's recursion theory \cite{Kleene} 
and von Neumann's random-access
stored-programs (RASP) \cite{RASP}, have further strengthened this view.

Nine decades have passed, yet it remains puzzling that the deep connection
between countability and computability has not been fully investigated.
The most related work is the theory of numberings that uses computable
mappings \cite{NUM}. This approach successfully introduced  
computability for the arithmetical hierarchy, which 
cover many uncomputable problems \cite{NUM2}.
Hack {\em et al.} also 
used computable bijections as a tool to exploit the computability beyond Turing machines
in the framework of ordered structures that cover uncountable sets \cite{HBG}. 
Our investigation is focused on the connection between countability and computability
over countable sets and has not attempted to introduce any new computing models
beyond Turing machines. 

To facilitate the presentation of our investigation, we first review some key concepts.
Following Cantor, a set $S$ is {\em countable} if either $S$ is finite or there exists
a bijection $f:\N\func S$, where $\N$ is the set of natural numbers \cite{Cantor}.
From now on, we refer to any such
bijection $f:\N\func S$ as a {\em counting bijection} of $S$. 
We say $S$ is {\em countably infinite} if it is infinite and countable.
It is easy to see that $S$ is countably infinite if and only if it
admits a counting bijection.
$S$ is said to be {\em enumerable} if it is either finite or admits a computable counting bijection. 
If $S$ is unenumerable, then every counting bijection of $S$ (if exists) is uncomputable.
$S$ is {\em increasingly countable} if $S\subseteq \N$ and 
it has an increasing counting bijection $f$ (i.e., $f(x) > f(y)$ whenever $x>y$).
A {\em characteristic function} of $S\subseteq\N$ is a total Boolean function
$f:\N\func \{0,1\}$ such that $S$ is equal to $\{ x\in\N \mid f(x) = 1\}$.
Since $f$ is total, all the characteristic functions of $S$ are equivalent.

Our investigation of the connection between countability 
and computability has focused on three main topics:

\begin{enumerate}
\item Our initial investigation shows that
a formal language $L$ is Turing recognizable if and only if $L$ is enumerable, and that
$L$ is decidable if and only if $L$ has an increasing computable counting bijection \cite{ZZ}.
This provides an alternative method for determining whether a formal language is
decidable, undecidable yet recognizable, or unrecognizable.

\item We introduce the concept of {\em nearby order} which requires that
the number of distinct elements lying between any two elements 
(when placed according to the order) be finite. A {\em counting order}
is a nearby well-order. We prove that a set $S$ is countable 
if and only if $S$ admits a counting order. 
Given a counting order $\toge$, one can design a procedure {\em deleteMin}$(S, \toge)$
that removes the minimal element from $S$. We show that
$S$ is countable if and only if such a procedure exists, 
and that $S$ is enumerable if and only if {\em deleteMin}$(S, \toge)$ is an algorithm.
This yields a necessary and sufficient procedural criterion for countability.

\item By {\em first-order arithmetic}, we mean a collection of 
the axiom systems in the language of first-order logic for the theory of natural numbers. 
{\em Peano arithmetic} is a typical example of first-order arithmetic \cite{Smullyan}.
Definability in first-order arithmetic has greater expressivity power than the axiom system
of {\em partial recursive functions}, which are equivalent to Turing machines. 
Given a set $S$ of natural numbers, we prove that
a counting bijection of $S$ is definable in first-order arithmetic
if and only if a characteristic function of $S$ is definable in first-order arithmetic. 
In the process of proving this result, we found that the standard proof of the claim
that every subset of $\N$ is countable is incorrect.
Although this claim is widely accepted in logic and mathematics,
this would imply that the set of all functions definable in first-order 
arithmetic is uncountable---a contradiction to the fact that 
only a countable number of functions can be definable in first-order arithmetic.
\end{enumerate}

This article is a substantial revision of \cite{Z1}. The next three sections
address the three topics outlined above, respectively.  
The common theme running through these topics is the relationship between countability
and computability. The first topic illustrates how the computability of a set
relates to the properties of its counting bijections.
The second topic offers a procedural respective on countable sets.
The final topic goes beyond Turing computability and demonstrates
how counting bijections and characteristic functions are related.

\section{Equivalence of Recognizable and Enumerable}

The main part of this section appears in the textbook \cite{ZZ} 
and is used here as background material for the entire article.
We assume familiarity with the basic concepts of theory of computation
\cite{Sipser,ZZ}. 

Given a Turing machine $M$, the language $L(M)$ 
{\em recognized} by $M$ is said to be {\em recognizable}. 
A set is {\em computable} if and only if it can be represented by a recognizable language. If $M$
halts on every input, then $L(M)$ is {\em decidable} and $M$ is a {\em
  decider}. A function $f:\N\func\N$, where $\N$ denotes the set of
natural numbers, is {\em computable} if $f$ can be computed by a
Turing machine $M$. The machine $M$ is said to be an {\em algorithm} if $M$
halts on every input; in this case, $f$ is {\em total computable}
(i.e., both total and computable). 

We need a computing model called {\em enumerator}, which is a
variant of Turing machine.  Informally, an enumerator is a Turing machine $M$ 
equipped with an attached printer \cite{Sipser}. The machine $M$ uses the printer as an
output device to print strings. The set $E(M)$ of printed strings is
the language {\em enumerated} by $M$.  The following result is known
for enumerators \cite{Sipser,ZZ}:

\begin{thm}\label{enumerator}
  $(a)$ If $L=E(M)$, where $M$ is an enumerator, then $L$ is recognizable. \\
  $(b)$ If $L$ is a recognizable formal language, there exists an
  enumerator $M$ such that $L=E(M)$ and every printed string by $M$ is
  unique.
\end{thm}

Let $g: \Sigma^*\func \N$ be a computable
bijection, where $\Sigma^*$ is the set of all finite-length strings over
the alphabet $\Sigma$.  For any $x,y\in\Sigma^*$,
we define $\toge$ over $\Sigma^*$ by $x \toge y$ if $g(x) \ge g(y)$.
The order $\toge$ is called the {\em canonical order} of $\Sigma^*$.
For any formal language $L\subseteq\Sigma^*$,  a bijection $f:\N\func L$ 
is {\em increasing} if $f(x)\toge f(y)$ whenever $x\ge y$ for any $x,y\in\N$.
For decidable languages, the following result is known \cite{Sipser,ZZ}.

\begin{thm}\label{decidable}
  An infinite formal language $L$ is decidable if and only if there exists an
  enumerator $M$ such that $L=E(M)$ and the printed strings are
  increasing in the canonical order.
\end{thm}

The following two propositions show that a formal language is
enumerable if and only if it is recognizable \cite{ZZ}.

\begin{prop}\label{main1}
  Every recognizable formal language is enumerable.
\end{prop}
\Proof Suppose $L$ is recognizable. The case when $L$ is finite is
trivial.  If $L$ is infinite, by Theorem\,\ref{enumerator}$(b)$, we
have an enumerator $M$ such that $L = E(M)$ and every printed string
by $M$ is unique, then $E(M)$ is countable because 
the order in which the strings are printed by $M$ defines a
computable bijection $f: \N \func E(M)$. That is, for $n\in \N$, we
compute $f(n)$ by algorithm $A(n)$ as follows:
\begin{quote} {\bf Algorithm} $A(n)$: Let $c:= 0$ and simulate $M$.
  When a string $x$ is printed by $M$, check if $c=n$. If yes, return
  $x$; otherwise, $c:= c+1$ and continue the simulation.
\end{quote}
Algorithm $A$ will terminate because $E(M)$ is infinite and $c$, which
records the number of printed strings by $M$, will reach $n$
eventually.  $A(n)$ can be modified as $B(w)$ to compute $f^{-1}(w)$
for $w\in E(M)$: Instead of checking $c=n$, check if $x=w$ and, if
yes, return $c$.  Since both $f$ and $f^{-1}$ are total functions, $f$
is a bijection.  Algorithm $A(n)$ is the evidence that $f$ is
computable.  \Endproof

\begin{prop}\label{main}
  Every enumerable formal language is recognizable.
\end{prop}
\Proof The case when formal language $L$ is finite is trivial.  If $L$
is enumerable, then there exists a computable bijection
$f: \N \func L$, Thus, we may use $f$ to design an enumerator $M$ that
prints strings $f(0)$, $f(1)$, $f(2)$, and so on.  It is evident that
$E(M) = L$. By Theorem~\ref{enumerator}$(a)$, $L$ is recognizable.
\Endproof

The following theorem combines the above two propositions into one and extends
the result from formal languages to general sets \cite{ZZ}.

\begin{thm}\label{setcount}
  An infinite set is enumerable if and only if it is computable.
\end{thm}
\Proof 
If a set $S$ is enumerable, then there exists a computable counting
bijection $f:\N \func S$.  Then $S = \{ f(n) \mid n\in\N\}$ 
because $f$ is a bijection.  $S$ is computable because
$f(n)$ is computable. 

When $S$ is unenumerable, there are two cases to consider:
$S$ can be represented by a formal language or not.  If yes, then $S$
is unrecognizable by Proposition\,\ref{main1}, thus uncomputable.  If not,
then $S$ cannot be recognized by any Turing machine, thus
uncomputable.  \Endproof

The above result is useful when we want to show some sets are
countable.  For instance, if we know a set is recognizable, then this
set is enumerable, hence countable. If a set is unrecognizable, then
it is unenumerable.  It is known that the collection of recognizable
sets is closed under intersection (i.e., the intersection of any two
recognizable sets is recognizable). It implies that the intersection
of two enumerable sets is enumerable, thus countable.

The following theorem is important about formal languages \cite{ZZ}.

\begin{thm}\label{main2}
  If $L$ is an infinite formal language, then
  \begin{enumerate}
  \item $L$ is decidable if and only if there exists an increasing and computable 
    bijection $f:\N\func L$.
  \item $L$ is recognizable if and only if there exists a computable bijection $f:\N\func L$.
  \item $L$ is unrecognizable if and only if every bijection $f:\N\func L$ is uncomputable.
  \end{enumerate}
\end{thm}
\Proof  For (1), by Theorem\,\ref{decidable}, $L$
is decidable if and only if the elements of $L$ can be enumerated in increasing
order. This enumeration defines an increasing and computable bijection
as in the proof of Proposition\,\ref{main1}, and an increasing and
computable bijection will produce an enumerator that enumerates
elements in increasing order as in the proof of Proposition\,\ref{main}.
Theorem~\ref{decidable} is then used to obtain the expected result.

 (2) comes from Theorem\,\ref{setcount}, as 
 ``$L$ is recognizable'' is equivalent to ``$L$ is enumerable'', and the latter
 means ``there exists a computable bijection $f:\N\func L$.''  
 
 (3) is logically equivalent to (2).
\Endproof

The above theorem holds if $L$ is a set of natural numbers and 
shows that the properties of counting bijection
$f:\N\func L$ is crucial to dictate whether $L$ is decidable or
computable. In particular, if $L$ is computable yet undecidable, then
its counting bijection must be computable but not increasing.

\begin{exmp}
 Consider the following encoding of some decision problems,
 where $\tuple{M}\in\Sigma^*$ is the string encoding Turing machine $M$ \cite{Sipser}.

\begin{itemize}

\item
  $E_{TM} = \{ \tuple{M} \mid L(M) = \E\,\mbox{\rm for Turing machine}~M \}$

\item
  $N_{TM} = \{ \tuple{M} \mid L(M) \neq \E\,\mbox{\rm for Turing machine}~M \}$

\item
  $All_{TM} = \{ \tuple{M} \mid L(M) = \Sigma^*\,\mbox{\rm for Turing machine}~M \}$
\end{itemize}

They are formal languages since $\tuple{M}\in\Sigma^*$.
$E_{TM}$, $N_{TM}$, and $All_{TM}$ are the encoding of all Turing
machines that accept, respectively, nothing, something, and
everything.  $N_{TM}$ is the complement of $E_{TM}$.  It is known that
$N_{TM}$ is recognizable and $E_{TM}$ is unrecognizable.  By
Proposition\,\ref{main1}, $N_{TM}$ is enumerable. 
By Proposition\,\ref{main}, $E_{TM}$
is unenumerable.  Both $All_{TM}$ and its complement,
$\ov{All_{TM}}$, are unrecognizable, hence, unenumerable.
\end{exmp}

\section{A Procedure for Countability}

In mathematics, a {\em poset} (partial order set) is a pair
$(S, \toge)$, where $S$ is a set and $\toge$ is a partial order of
$S$---that is, $\toge$ is an antisymmetric, transitive, and reflexive
relation on $S$.  For any $a,b\in S$, we write
$a \succ b$ to mean $a\toge b$ and $a \neq b$. 
The order $\toge$ is {\em total} if, for any $a,b\in S$, exactly one of $a \succ b$,
$b\succ a$, or $a=b$ holds.  It is {\em well-founded} if every non-empty
subset of $S$ has a minimal element with respect to $\toge$.  
The order $\toge$ is a {\em
  well-order} if it is both total and well-founded.  Finally, $\toge$ is {\em
  dense} if, for any $a,b\in S$ with $a \succ b$, there exists an element $c\in S$
such that $a\succ c\succ b$.

\begin{defn} Given a poset $(S, \toge)$, 
   $\toge$ is said to be {\em nearby} if, for any two elements
  $a, b\in S$, the set $\{ c \in S \mid a \succ c \succ b\}$
  is finite;
   $\toge$ is said to be a {\em counting order} of
  $S$ if $\toge$ is a nearby well-order.
\end{defn}

\begin{exmp}
Here are a few examples: $\ge$ is a counting order of $\N$; $\ge$ is not a
counting order for $\Z$, the set of integers, because $\ge$ is not a
well-order of $\Z$.

Let $\ge$ be the order on $\N\times\N$ such that
$(x_1, y_1) \ge (x_2, y_2)$ if $x_1 > x_2$ or
$x_1=x_2$ and $y_1\ge y_2$ (that is, $\ge$ is a lexicographic order of
$\N\times\N$). $\ge$ is a well-order of $\N\times\N$.  However,
$\ge$ is not a counting order because it is not nearby.  There
are infinite pairs between $(0, 0)$ and $(1, 0)$: 
the set $\{ (0, n+1) \mid n\in\N\}$ is infinite.

Note that a dense order is never nearby, and 
the concepts of ``nearby'' and ``non-dense'' are different: 
$\ge$ on $\N\times\N$ is neither dense nor nearby.
\end{exmp}

When proving that the set of reals is not countable, 
Cantor's diagonal argument is used to show that the assumed counting order is
not nearby because some elements are left out of the order.
The following result shows the connection between counting order
and countability.

\begin{thm}\label{orderc}
A set is countable if and only if it admits a counting order.
\end{thm}
\Proof The finite case is trivial, so we consider only infinite sets.
If $S$ is countable, then $S$ has a counting bijection $f:\N \func S$.
Define $\toge$ over $S$ by $f(x) \toge f(y)$ if $x \ge y$ for any $x,y\in\N$.
Then $\toge$ is a nearby well-order of $S$, since $\ge$ is a nearby well-order of $\N$.

On the other hand, let $\toge$ be a nearby well-order of $S$.  
For any $n\in\N$, define $g(n)$ to be the $(n+1)^{th}$ minimal element of $S$
under $\succ$. $g:\N\func S$ is injective because $i^{th}$ minimal element
is different from $j^{th}$ minimal element if $i \neq j$. 
$g$ is surjective because $\toge$ is a counting order. That is,
let $a = min(S)$, $b\in S$ and $a\neq b$. Since $\toge$ is nearby,  
the set $X = \{ c\in S \mid b \succ c \succ a \}$ is finite.
If $|X| = n$, then $b$ is the $(n+2)^{th}$ minimal element of $S$. Hence,
for any $b\in S$, there exists $m\in\N$, $g(m) = b$. Thus,
$g$ is a counting bijection of $S$.
\Endproof

If $S$ is computable yet undecidable, by Theorem~\ref{main2},
we can obtain a computable bijection $f:\N \func S$, where $f$ is not increasing
(with respect to $\ge$).
In the above proof, from $f$, we obtain a counting order $\toge$ of $S$
such that $f$ is increasing with respect to $\toge$. Obviously, $\toge$ is different
from $\ge$. 

In the second part of the above proof, we defined the function
$g(n)$ to be the $(n+1)^{th}$ minimal element of $S$
under $\succ$.  The same idea was used in the proof of the claim
that every set of natural numbers is countable \cite{Tao}.
If $S$ is undecidable, then $g$ cannot be computable if $\succ$ is $>$ over $\N$,
because an increasing computable $g$ implies that $S$ is decidable.

Theorem~\ref{orderc} claims that a set has a counting bijection if and only if it has a counting order. This
provides a necessary and sufficient condition for countably infinite sets.

Given a poset $(S, \toge)$, where $\toge$ is a counting order of $S$, 
we will design a procedure {\em deleteMin$(S, \toge)$} 
that removes the minimal element (with respect to $\toge$)
from $S$ and returns the minimal element.
This procedure offers a new perspective for studying countability.  

\begin{thm}\label{delete}
\begin{enumerate} 
\item $S$ is countable if and only if procedure deleteMin$(S, \toge)$ exists.
\item $S$ is enumerable if and only if there exists an algorithm deleteMin$(S, \toge)$.
\end{enumerate}
\end{thm}
\Proof In the following proof, we consider only infinite sets as the finite case is trivial.

$(1)$ By Theorem~\ref{orderc}, $S$ is countable if and only if $S$ has a counting
order. Procedure {\em deleteMin}$(S, \toge)$ exists 
if and only if $\toge$ exists, because for nonempty $S$,
\[min(S, \toge) = x \IFF (x \in S \AND \forall y\in S~\neg(x \succ y)).\]
We may design {\em deleteMin}$(S,\toge)$ which scans $S$
once to locate $min(S,\toge)$, deletes it from $S$ and returns it.

$(2)$  If $S$ is enumerable, there exists a computable bijection $f: \N\func S$
and $S = \{ f(n) \mid n\in\N\}$. In fact, $f$ is total computable since $f$ is total.
Thus, the Turing machine that computes $f$ is an algorithm.
As in the proof of Theorem~\ref{orderc},  for any $x,y\in\N$, we define $\toge$ by
$f(x) \toge f(y)$  whenever $x \ge y$. Then $min(S, \toge) = f(0)$ 
and the successive calls of {\em deleteMin}$(S,\toge)$
will return $f(0)$, $f(1)$, $f(2)$, .... Since $f$ is total computable, 
so is {\em deleteMin}$(S,\toge)$.
On the other hand, if {\em deleteMin}$(S,\toge)$ is an algorithm,  for any $n\in\N$,
let $f(n)$ be the element returned by the $(n+1)^{th}$ call of {\em deleteMin}$(S, \toge)$,
then $f$ is a computable counting bijection of $S$.
\Endproof

Theorem\,\ref{delete} gives us a necessary and sufficient condition for an
enumerable set $S$: Having an algorithm {\em deleteMin} with a
counting order $\toge$.  Using algorithm {\em deleteMin}, 
every element of $S$ can be enumerated one by
one from the smallest by the counting order $\toge$. 
If $S$ is a formal language, then the enumerated numbers are not
necessarily increasing in the canonical order. For instance, let $H$ be the 
formal language for the halting problem of Turing machines. 
Since $H$ is computable yet undecidable, $H$ cannot be
increasingly enumerated in the canonical order (see Theorem\,\ref{decidable}).

\section{Unproven Statements about Countable Sets}
\label{open}

The claim that ``every subset of $\N$ is countable'' 
is widely accepted and taught in many college courses. 
The claim has several equivalent statements.

\begin{prop}\label{four} The following statements are logically equivalent:
\begin{enumerate}
\item Any subset of a countable set is countable.
\item Any subset of $\N$ is countable.
\item If there is an injective function from set $S$ to $\N$, then $S$ is
  countable.
\item If there is a surjective function from $\N$ to $S$, then $S$ is
  countable.
\end{enumerate}
\end{prop}

\Proof
$(1)\Rew (2)$: $\N$ is countable.

$(2)\Rew (3)$: Let $f: S\func \N$ be injective and
$X = \{ f(x) \mid x \in S\}$. Since $X\subseteq \N$, $X$ is countable
by $(2)$.  Note that $f: S\func X$ is a bijection from $S$ to $X$,
that is, $S$ and $X$ are the same size. So, $S$ must be countable.

$(3)\Rew (4)$: Let $g: \N\func S$ be surjective, then
for every $x\in S$, the set $g^{-1}(x) = \{ y\mid g(y) = x\}$
is not empty. Define $f: S\func \N$ by $f(x) = min(g^{-1}(x))$,
then $f$ is injective. By $(3)$, $S$ is countable.

$(4)\Rew (1)$: Let $X\subseteq S$ and $S$ be countable.  If $X$ is
finite, then $X$ is countable.  If $X$ is infinite, then
$S$ must be infinite and there exists a bijection $h: \N\func S$. Let
$a$ be an element of $X$.  Define $g: \N\func X$ by $g(x) = h(x)$ if
$h(x)\in X$, otherwise $g(x)=a$. Then $f$ is surjective. By
$(4)$, $S$ is countable. If $g$ is not required to be total, then
``$g(x)=a$'' is not needed when $h(x)\notin X$.  \Endproof

If these statements are true, we may have an alternative yet simpler definition of
``countable'': A set $S$ is {\em countable} if there is an injective
function from $S$ to $\N$.  Some textbooks use this definition
\cite{EndertonC}.  This definition does not matter whether $S$ is finite or 
not and uses only injection, not bijection. 

These statements appear in many
textbooks on logic \cite{Enderton}, set theory \cite{Goldrei,Hrbacek}, 
discrete mathematics \cite{dis},
or theory of computation \cite{Martin}.
In a textbook on set theory \cite{EndertonC}
published in 1977, the author wrote without proof that
``obviously every subset of a countable set is countable.''  
It also appears in an influential textbook by Terence
Tao \cite{Tao} ({Proposition}\,8.1.5).  
\\[1.5pt]

\noindent {\bf Proposition 8.1.5} of \cite{Tao} {\em Let $X$ be an
  infinite subset of the natural numbers $\N$. Then there exists a
  unique bijection $f: \N\func X$ which is increasing, in the sense
  that $f(n + 1) > f(n)$ for all $n \in \N$. In particular, $X$ has
  equal cardinality with $\N$ and is hence countable.}
\\[1.5pt]

Proposition 8.1.5 claims that 
every subset of $\N$ is increasingly countable, which is stronger
than the claim that  every subset of $\N$ is countable.
We show later that the two claims are equivalent in first-order arithmetic.
Unfortunately,  Proposition 8.1.5 lacks a valid proof because its proof 
leads to a contradiction. 
In fact, none of the four statements in Proposition~\ref{four} has a valid proof.
Hence, their validity remains an open problem.
 
In the following, we show what went wrong in the proof of Proposition 8.1.5 
by starting with a formal definition of ``definable''. By convention,
an axiom system $L$ is a language of logic which uses an alphabet to construct
formulas and define inference rules. Some of these formulas are designated to be {\em axioms}
and some of these formulas are {\em theorems} which are deduced 
from the axioms and the inference rules. The power of an axiom system
is measured by its set of theorems.

Given an axiom system $L$,  
the semantics of syntactic items in $L$ is specified in a metalanguage $M$ which defines truths
of the formulas in $L$. After Tarski's approach on undefinability, 
people often use an algebraic structure, called {\em model} of $L$,
to describe the semantics of $L$ \cite{Tarski}. 
For instance, the structure of natural numbers is used as the standard model of first-order arithmetic.
This model is used to define {\em true arithmetic} $T$, the set of all sentences 
(i.e., closed formulas) of $L$ which are true in the model.
This model consists of $\N$ (called the {\em underlying set}), 
a collection of arithmetic operations on $\N$, 
and a set of {\em axioms} that these operations must satisfy.
These operations are presented by relations or functions over the underlying set..
Informally, a relation (or a function) of the model is said to be ``definable'' if
the relation can be specified by a formula of $L$.

We assume that in any axiom system $L$ of first-order arithmetic,
a natural number $n$ is represented
by the {\em numeral} $\ov{n} = s^n(0)$, where $0$ and $s$ (the unary {\em successor}
function) are the symbols of $L$.

\begin{defn}\label{relation}
For any axiom system $L$ using $0$ and $s(x)$, 
a relation $p(n_1, n_2, ..., n_k)$ in a model of $L$, where $n_i\in\N$ for $1\le i\le k$,
is said to be {\em definable} in $L$ 
if there exists a formula $\phi(x_1, x_2, ..., x_k)$ of $L$ 
such that $p(n_1, n_2, ..., n_k)$ is true if and only if  $\phi(\ov{n_1}, \ov{n_2}, ..., \ov{n_k})$ is true. 
A set $S\subseteq \N$ is said to be {\em definable} in $L$
if $S = \{ n\in\N\mid p(n) \}$, where $p(n)$ is a unary relation definable in $L$.
\end{defn}

For instance, let $F$ be the axiom system for {\em partial recursive functions}
 (no universal quantifiers in any formula).
According to the Church-Turing thesis, a set is definable in $F$ if and only if it is computable.
Conventional operations appearing in the standard model of natural numbers
are all definable in first-order arithmetic (both quantifiers are allowed), 
as their definitions can be expressed by the formulas of first-order arithmetic. 

Since a function is a special case of relation and a relation is often regarded
as a Boolean function, the above definition applies to functions and we will
use functions as a synonym of relation.
We will use the following proposition later.

\begin{prop}\label{limit}
For any axiom system $L$, if a model of $L$ uses $\N$ as its underlying set, 
then the set of functions of the model definable in $L$ is countable.
\end{prop}
\Proof Since $\N$ is countable, the alphabet of $L$ must countable, because
$\N$ cannot serve as the underlying set for an uncountable alphabet.
Thus, the set of formulas of $L$ is countable, because each formula uses
a finite set of symbols. Hence, the set of functions defined
by these formulas is countable.
\Endproof

If a set is definable in an axiom system $L$, it does not imply that 
its characteristic function is definable in $L$. That is,
if $S = \{ n\in\N\mid p(n) \}$, the unary relation $p(n)$ is not necessarily
a characteristic function of $S$, unless $p(n)$ is total.
For instance, if $S = \{ n\in\N\mid p(n) \}$ is decidable, then $p(n)$ is its characteristic function.
If $S$ is computable yet undecidable, then $p(n)$ is not its characteristic function in $F$,
since $p(n)$ is not total in $F$.
On the other hand, if the characteristic function of $S$ is definable in $L$,
$S$ is definable in $L$.


A proof sketch of Proposition 8.1.5 goes as follows \cite{Tao}:
Since the standard order $<$ is a well-order of $X\subseteq\N$, $X$ has a minimal element under $<$,
say $a_0 = min(X)$. Let
$X_0 = X$ and, for $i>0$, $X_i = X_{i-1} \ssub \{ a_{i-1} \}$, where
$a_i = min(X_i)$, we obtain an increasing sequence 
\[ a_0 < a_1 < a_2 < ...\] along with the sequence $X_0, X_1, X_2, ...$.
Define $f:\N\func X$ by $f(n) = a_n = min(X_n)$, it is evident that $f$ is an 
increasing bijection (see the proof of Theorem~\ref{orderc}). 
Hence, $X$ is increasingly countable.

\begin{prop}\label{firsto}
The increasing counting bijection constructed in
the proof of Proposition 8.1.5 is definable in first-order arithmetic.
\end{prop}
\Proof At first, the function $min(X)$ can be defined in first-order arithmetic:
\[(min(X) = a) \IFF (a \in X \AND \forall b\in X~\neg(a > b)).\]
Secondly, the counting bijection $f:\N\func X$ in the proof can be expressed
by the function $g$ defined as follows: For all $x\in\N$, $f(n) = g(X, n)$, where
\[\begin{array}{rcl}
g(X, 0) & = & min(X) \\
g(X, n+1) & = & g(X - \{ min(X) \}, n)
\end{array}\]
Assuming $X_0 = X$ and, for $i>0$, $X_i = X_{i-1} \ssub \{ a_{i-1} \}$, where
$a_i = min(X_i)$.
It is straightforward to check that 
\[f(n) = g(X_0, n) = g(X_1, n-1) = g(X_2, n-2) = \cdots = g(X_n, 0) = min(X_n) = a_n.\] 
Hence, the proof of Proposition 8.1.5 can be carried out in first-order arithmetic.
\Endproof

The following result presents a close connection between 
the definability of sets and their counting bijections in first-order arithmetic.

\begin{thm}\label{strongest} Let $S$ be an infinite set of natural numbers.
The following statements are equivalent in first-order arithmetic.

\begin{enumerate}
\item An increasing counting bijection of $S$ is definable in first-order arithmetic.

\item A counting bijection of $S$ is definable in first-order arithmetic.

\item A characteristic function of $S$ is definable in first-order arithmetic.
\end{enumerate}
\end{thm}
\Proof $(1)\Rew (2)$: $(1)$ is a special case of $(2)$.

$(2)\Rew (3)$:
Let $f:\N\func S$ be a counting bijection of $S$. For any $n\in\N$, let
$p(n)$ be $\exists x~(f(x) = n)$. Then $n\in S$ if and only if $p(n)$ is true.
If $f$ is definable in first-order arithmetic, so is $p$.
Since $f$ is total, $p(n)$ is total, too. Hence, $p(n)$ is a characteristic function of $S$.

$(3)\Rew (1)$: 
Suppose $p(n)$ is a characteristic function of $S$ that is definable in first-order arithmetic.
For any $n\in\N$, define $f(n) = g(n, 0)$, 
where $g: \N\times\N \func \N$ is defined as follows:
\[\begin{array}{rcl}
          g(0, m) & = & \IF p(m)  ~\THEN  m~ \ELSE  g(0, m+1)\\
      g(n+1, m) & = & \IF p(m) ~\THEN  g(n, m+1) ~ \ELSE  g(n+1, m+1)
\end{array}\] 
Let $S = \{ a_0, a_1, a_2, ... \}\subseteq\N$ such that $a_i < a_{i+1}$ for $i\in\N$.
We show below that $f(n) = a_n$ by induction on $n$. For the base case,
$f(0)=g(0,0)$. The recursive calls of $g(0,0)$ will return the least $m$ such that
$p(m)$ is true. Obviously, $m=a_0$. 

Now we consider the inductive case.
For any $n\in\N$, $g(n+1,0)$ visits $0, 1, ..., a_0$ (as the second parameter of $g$) 
before the first parameter of $g$ is decreased by $1$
in the recursive calls of $g$. Similarly, $g(n, a_0+1)$
will visit $a_0+1, ..., a_1$; $g(n-1, a_1+1)$ will visit $a_1+1, ..., a_2$, and so on,
during the recursive calls. Finally, $g(0, a_{n}+1)$ will return $a_{n+1}$, 
the minimal number of $S - \{f(0), f(1), ..., f(n)\}$. 
It is ready to check that 
\[f(n+1) = g(n+1, 0) = g(n, a_0+1) = g(n-1, a_1+1) = \cdots = g(0, a_{n}+1) =
a_{n+1}. \]
It is easy to check that $f$ is an increasing counting bijection of $S$
and $f$ is definable in first-order arithmetic.
Note that $g(n,m)$ is total because
$S$ is infinite and $p(m)$ is true for an infinite number of $m$. 
\Endproof

The above theorem shows the coexistence of a counting bijection
and a characteristic function for an infinite set of natural numbers.
That is, if one exists, so does the other; if one does not exist, neither does the other.

The {\em arithmetical hierarchy} 
classifies formulas of first-order arithmetic by the use
of (unbounded) quantifiers \cite{Soare}.  The classifications of
formulas are denoted by $\Sigma^0_{n}$ (and $\Pi^0_{n}$) for $n\in\N$
($0$ in the superscript denotes first-order). 
For $n>0$, a $\Sigma^0_{n}$ formula is equivalent to a prenex formula
that begins with some existential quantifiers and alternates $n-1$
times between series of existential and universal quantifiers.
Analogously, a $\Pi^0_{n}$ formula, where $n>0$, is equivalent to a prenex formula
that begins with some universal quantifiers and then alternates.
A set of natural numbers defined by a formula of $\Sigma^0_n$ ($\Pi^0_n$)
is called a $\Sigma^0_n$-{\em set} ($\Pi^0_n$-{\em set}).
It is known that each $\Sigma^0_0$-set is decidable;
each $\Sigma^0_1$-set is (partial) computable;
each $\Pi^0_1$-set is co-computable. For $n>1$,
a $\Sigma^0_n$-set may be uncomputable  \cite{Soare}.
Since the arithmetical hierarchy provides an infinite sequence of collections of
problems with increasing complexity,
it has been viewed as a test bed for computability beyond Turing machines.
The expressiveness powers of the arithmetical hierarchy and first-order arithmetic
are equivalent.
Theorem~\ref{strongest} shows for the first time the equivalence
of  first-order arithmetical sets and the countable sets definable in first-order arithmetic. 
It also displays the connection between the existence of counting bijections
and computability beyond Turing machines.

Since the power set of $\N$ is uncountable, by Proposition~\ref{limit},
 there exist uncountably many subsets of $\N$ that 
 are undefinable in first-order arithmetic, because each distinct subset of $\N$
 needs a distinct Boolean function of first-order arithmetic for its definition. 
 If an infinite $S\subseteq\N$ is undefinable in first-order arithmetic,
 so is the characteristic function of $S$. By Theorem~\ref{strongest}, any counting 
 bijection of $S$ is undefinable in first-order arithmetic, which contradicts
 the claim that every subset of $\N$ is countable.
 
\ignore{
 Proposition 8.1.5 of \cite{Tao}
 we should accept the claim
that $S$ has a characteristic function in first-order arithmetic, which would imply 
a contradictory claim that every infinite $S\subseteq\N$ is definable in first-order arithmetic.
}

\begin{prop}\label{final}
Proposition 8.1.5 of \cite{Tao} lacks a valid proof.
\end{prop}
\Proof By Proposition~\ref{limit}, there
are uncountably many subsets of $\N$ which are undefinable
in first-order arithmetic, and so are their characteristic functions and counting bijections. 
On the other hand, the proof of Proposition 8.1.5 defines a counting bijection
for every infinite set $X\subseteq\N$. By Proposition~\ref{firsto},  using $X$ as a parameter,
this definition can be expressed by a single formula of first-order arithmetic (no
distinct formula for each distinct set). Apparently, there exists a contradiction
when $X$ is used as parameter.

What is the role of using $X$ as parameter? 
The variable $X$ denotes only its existence as a subset of $\N$. However, it does not imply
the existence of its characteristic function. $X$ often appears together with
the operator $\in$, such as in formula $a\in X$. 
For any axiom system $L$ of first-order arithmetic, where $\N$ is the underlying set in its model,
the interpretation of $\in$ in the formula $a\in X$ is the characteristic function of $X$ of the model for $L$.
We need this function to tell if an element is a member of $X$ or not in the model. 

The proof of Proposition~8.1.5 implicitly assumed that the characteristic function exists for every $X\subset\N$.
For instance, the proof uses the definition of $min(X)$, 
\[min(X) = a \IFF (a \in X \AND \forall b\in X~\neg(a > b)),\]
and the interpretations of $a\in X$ and $b\in X$ need the characteristic functions of $X$.
If the characteristic function of $X$ is undefinable in first-order arithmetic,  then
the implicit assumption is false.  For the validity of the proof, we need
the condition that ``the characteristic function of $X$ is definable in first-order arithmetic.'' 
Proposition 8.1.5 simply ignored this condition, an error of hasty generalization.
Hence, the proof of Proposition 8.1.5 cannot be valid as it leads to the contradiction. 
\Endproof

\begin{exmp}\label{Tstar}
A notable example of undefinable sets comes from Tarski's undefinability theorem.
Let $T^*$ denote the set of \Godel\ numbers of all true sentences 
in the standard model of first-order arithmetic. The following three propositional variables
have no consistent interpretations:

\begin{itemize}
\item $p$ means ``$T^*$ is definable in first-order arithmetic.'' It is well-known that
$p$ is false \cite{Tarski,Smullyan}.

\item $q$ means ``$T^*$ has a counting bijection definable in first-order arithmetic.''
By Proposition 8.1.5 \cite{Tao}, $T^*\subset\N$ has a
counting bijection definable in first-order arithmetic (Proposition~\ref{firsto}).
Hence, $q$ must be true.
 
\item $r$ means ``the characteristic function of $T^*$ is definable
in first-order arithmetic.'' If $r$ is true, then $p$ must be true because
the characteristic function of $T^*$ provides a definition of $T^*$.
If $r$ is false, then $q$ must be false by Theorem~\ref{strongest}.
\end{itemize}
In short, we cannot find a consistent interpretation for $\{ p, q, r\}$.
\end{exmp}

\section{Conclusions}

In Cantor's time, the notion of {\em computability} had not yet been
formally defined.  
Cantor used the word {\em count} to indicate that an operation is performed. 
It turns out that the properties of counting bijections are closely 
related to computability.
Our initial investigation shows that 
a set is computable if and only if it has a computable counting bijection,
and that a set is decidable if and only if it has an increasing 
and computable counting bijection \cite{ZZ}.
In this article, we provide a procedural perspective on countability.
We show that a set $S$ is countable if and only there exists
a procedure {\em deleteMin}$(S, \toge)$ that deletes
the minimal element of $S$ according to $\toge$, a counting order of $S$.
Moreover, $S$ is computable if and only if {\em deleteMin}$(S, \toge)$ is
an algorithm. This result further strengthens the relationship between
countability and computability.

For computability beyond Turing machines, one may use first-order
arithmetic as a specification tool. We show that, for an infinite set $S$ 
of natural numbers, a characteristic function of $S$ is definable in first-order arithmetic
if and only if a counting bijection of $S$ is definable in first-order arithmetic.
In the process of proving this result, we found that the standard proof
of the claim that every subset of $\N$ is countable is invalid, as
the proof of the claim leads to the conclusion that the set of counting
bijections definable in first-order arithmetic is uncountably many,
which contradicts the fact that only a countable number of counting bijections
is definable in first-order arithmetic.

Without a valid proof, the claim remains an open problem.
Proving it appears to be very challenging.
By Proposition~\ref{limit},  the claim cannot be proved within any axiom
system whose model uses $\N$ as the underlying set.
We may need an axiom system that does not use $\N$ as its underlying set.
It is unclear how $\N$ and the counting bijections of $S\subseteq\N$
are defined in this axiom system.
According to \Godel's incompleteness theorems \cite{Godel}, every 
sufficiently powerful axiom system has inherent limitations. 
These limitations may introduce a set of natural numbers whose countability cannot
be established in the same axiom system.
There appears to be no upper
bound on the difficulty or complexity in either computability or countability.

\end{document}